\documentclass[journal]{IEEEtran}
\usepackage{amsmath,amsfonts}
\usepackage{float}
\usepackage[ruled,vlined]{algorithm2e}

\usepackage{graphicx}
\usepackage{array}
\usepackage{type1cm}
\usepackage{lettrine}
\usepackage{amssymb}
\usepackage{color}
\usepackage{hyperref}
\usepackage[letterpaper, left=0.62in, right=0.62in, bottom=1in, top=0.701in]{geometry}

\usepackage{tabularx}
\usepackage{balance}

\begin{document}
\bstctlcite{IEEEexample:BSTcontrol}
\title{ Multi-Agent Reinforcement Learning for Network Selection and Resource Allocation in Heterogeneous multi-RAT Networks}
\author{Mhd Saria Allahham, \textit{Student Member, IEEE}, Alaa Awad Abdellatif, \textit{Member, IEEE} , Naram Mhaisen,  Amr Mohamed, \textit{Senior Member, IEEE}, Aiman Erbad, \textit{Senior Member, IEEE} and Mohsen Guizani, \textit{Fellow, IEEE}
\thanks{Mhd Saria Allahham is now with School of Computing, Queen's University, Canada. This wok was done while he was a research fellow at Qatar University (e-mail: ma1517219@qu.edu.qa).}
\thanks{Naram Mhaisen is now with the Faculty of Electrical Engineering, Mathematics and Computer Science, Delft University of Technology, Netherlands. This wok was done while he was a research fellow at Qatar University (e-mail: nm1300940@qu.edu.qa).}
\thanks{Alaa Awad Abdellatif and Amr Mohamed are with College of Engineering, Qatar University, Qatar (e-mail: alaa.abdellatif@ieee.org, amrm@qu.edu.qa).}
\thanks{Aiman Erbad is with Division of Information and Computing Technology, College
of Science and Engineering, Hamad Bin Khlifa University, Qatar (e-mail: aerbad@hbku.edu.qa).}
\thanks{Mohsen Guizani is with Machine Learning Department, MBZUAI, Abu Dhabi,
UAE. (e-mail: mguizani@gmail.com).}
}

\maketitle
\begin{abstract}
The \color{black} {rapid production of mobile devices along with the wireless applications boom is continuing to evolve daily. This motivates the exploitation of wireless spectrum using multiple Radio Access Technologies (multi-RAT) and developing innovative network selection techniques to cope with such intensive demand while improving Quality of Service (QoS).    
%
Thus, we propose a distributed framework for dynamic network selection at the edge level, and resource allocation at the Radio Access Network (RAN) level, while taking into consideration diverse applications' characteristics. In particular, our framework employs a deep Multi-Agent Reinforcement Learning (DMARL) algorithm, that aims to maximize the edge nodes' quality of experience while extending the battery lifetime of the nodes and leveraging adaptive compression schemes.  
Indeed, our framework enables data transfer from the network's edge nodes, with multi-RAT capabilities, to the cloud in a cost and energy-efficient manner, while maintaining QoS requirements of different supported applications. }  
Our results depict that our solution outperforms state-of-the-art techniques of network selection in terms of energy consumption, latency, and cost.   

\end{abstract}
\begin{IEEEkeywords}
Heterogeneous networks, edge computing, wireless healthcare systems, multi-RAT architecture, deep reinforcement learning. 
\end{IEEEkeywords}

\section{Introduction}
With the emerging of the Internet of Mobile Things (IoMT) and unprecedented 5G applications, several strict communication requirements have arisen. Such requirements demand wireless networks to be responsive while adapting to different applications' requirements. Specifically, for e-health applications, the ability of the healthcare systems to predict and instantly react to emergency situations is mandatory to reduce the risks of chronic diseases and the mortality rate \cite{ssHealth}. However, the strict latency requirements for emergency conditions along with the enormous amount of generated data are still major challenges for e-health systems. Such demand for ultra-low latency and other Quality of Service (QoS) requirements has motivated us to leverage the evolution of the 5G network toward dense Heterogeneous networks (HetNets) \cite{DynamicPricing}. 
5G HetNets are expected to enhance users' QoS, enable diverse performance improvements, and fulfill various service requirements through increasing the opportunity of spatial resource reuse  \cite{5GHetNet, 6GOpening}.
Indeed, leveraging multi-Radio Access Technology (RAT) will enable a device/edge (e.g., mobile phones, edge gateways) to utilize the available radio resources across various spectral bands to connect with the network infrastructure. Hence, the edge devices that are equipped with multiple interfaces (e.g., Wi-Fi, 3G, 4G, Bluetooth) will be able to simultaneously access the available networks with different RATs. However, this calls for designing innovative and scalable networks selection schemes that consider e-health QoS requirements while maintaining high spectrum efficiency across diverse networks. 

Artificial Intelligence (AI) has been a key feature in 5G networks recently, where introducing AI into HetNets can help in developing and executing efficient and intelligent network selection schemes \cite{5G_wang2015artificial_intro}. Moreover, in AI-based user-centric network selection, different users can have customized and dynamic selection behaviors based on their own needs and requirements for different applications \cite{5G_wang2019intelligent_intro}. Specifically, the most efficient dynamic network selection techniques are based on Reinforcement Learning (RL). RL is an emerging field in AI that studies optimal sequential decision making for an agent in non-deterministic environments \cite{sutton2018reinforcement}. In single-agent RL, the agent tries to learn the optimal policy from the interactions with the environment, aiming to maximize its reward. Nevertheless, in many environments of the real world, including self-driving cars, packet delivery, and others, there are multiple agents acting simultaneously on the environment, based on their own observation, with the aim of maximizing their own utility, which does not necessarily align with others. Single-agent RL modeling of such large-scale environments is not efficient, where the agent if deployed in a centralized manner (i.e, at the core network ) it will lack the scalability, since it has to have complete knowledge about the environment including users' actions.

As such, the modeling framework that captures such complex interactions between multi-agents is game theory. However, due to the difficulty of classical solution methods for games, especially for environments with many agents, data-driven solutions based on reinforcement learning have been emerging as a promising solution known as Multi-Agent RL (MARL). MARL extends the RL framework to explicitly model the existence of multiple agents and the effect of their joint action on the environment. In this framework, the solution concept is to reach a set of policies (a policy for each agent) that form an equilibrium with the maximum reward (i.e., set of dominant policies). Of course, reaching this equilibrium, if it exists, is more challenging compared to single-agent RL. This is primarily due to the non-stationarity nature of the environment (i.e., for an agent, the same policy might result in a different performance according to what other agents are doing, which is not necessarily known by that agent). 

MARL includes different sub-problems ranging from learning the cooperation between agents \cite{tan_marl, Fischer}, to learning the communication between different types of agents \cite{Foerster}. The recent advancements in deep learning along with its integration with the MARL lead to the new Deep Multi-agent Reinforcement Learning (DMARL) concept \cite{crandall2005learning,hausknecht2015deep}, where DMARL has been widely utilized for various tasks in 5G networks such as distributed resource allocation  \cite{gundougan2020distributed,guo2020joint} and interference management \cite{challita2018cellular}. Moreover, in edge computing, DMARL is being employed for numerous tasks such as computation offloading \cite{offloading_marl}, resource allocation and cashing \cite{aloocation_marl,cashing_MARL} and control of network traffic \cite{control_flow}. Overall, there has been impressive progress at the algorithmic level in the DMARL community, which is yet to be customized, adapted, and utilized in practical scenarios. In this work, we utilize and adapt state-of-the-art DMARL algorithms to jointly tackle a practical and important issue of network selection and resource allocation.






\section{Related work}\label{sec:related}
The related work in the literature exploits different methodologies for solving the network selection problem, including: game theory, Markov decision processes (MDPs), multi-attribute decision making, and optimization techniques \cite{consumer_dynamic}.  
Game theory approaches for network selection assume that the users aims to increase their rewards, while networks operators are trying to increase their revenues by increasing their number of users. However, it cannot be always guaranteed that the networks operators will act in a rational manner and the users will get the targeted QoS \cite{GAME_HETNET}, \cite{CASHING_AP}.   
Moreover, the proposed solutions, using game theory, are typically complexity-prohibitive, and their convergence to the optimal solution is not guaranteed. Even if they converge, it is not always guaranteed that they can converge to an optimal solution \cite{consumer_dynamic}. 
In other works, the network selection problem has been formulated as an MDP to investigate network switching between different RATs \cite{mdp_hetnet_wireless}, \cite{rat_select_cellular}.  However, obtaining an optimal solution using such approaches is again computationally intensive, especially in the case of large networks \cite{load_hetnet}. 

Network selection has been also studied using multi-attribute decision making. Such approaches assigned different weights for different factors affecting the network selection decision. Then, various techniques such as simple additive weighting \cite{drissi2016fuzzy}, multiplicative exponent weighting \cite{Comparison_dec_hetnet}, \cite{load_multimedia}, and grey relational analysis \cite{selection_hyprid},  \cite{integrated_UMTS}, have been considered for selecting the appropriate network. 
However, it is usually hard to prove that such approaches can obtain an optimal solution.   
Optimization techniques have been also investigated for solving network selection problem. Despite the guaranty of optimality using such approaches, typically, formulating the network selection problem as an optimization problem with reasonable complexity (to be run at the edge) is not an easy task. Obtaining the optimal solution, subject to different networks, applications, and power constraints may lead to an NP-hard problem \cite{Association_hetnet}.  Moreover, since network resources and characteristics can vary with time, as well as the edge resources, classical optimization approaches can be computationally expensive, where online optimization approaches such RL can be more appropriate.

RL-driven network selection has been used and studied in \cite{reinforcement_NS2_dynamics,reinforcement_NS,zina_RL_NS}, where the authors showed that RL can achieve faster convergence and optimal behaviors. However, the existence of multiple users that employ network selection schemes along with resource-constrained RANs calls for the need for MARL and distributed optimization to decentralize the individual policies for individual users and RANs, which would scale efficiently in large scale systems. The authors in \cite{MARL_NS2_smart} have proposed a strategy for multi-RAT access based on MARL with the aim of maximizing the average system throughput while satisfying the users' QoS preferences. However, the authors did not address the mobile devices characteristics (e.g., energy budget) and the application requirements (e.g., the quality of the transmitted medical data). 


Different from the aforementioned work in the literature, we aim to leverage the intelligence at the edge for optimizing networks selection decisions in the ultra-dense heterogeneous networks, and at the network side for optimizing the resource allocation for the edges. 
Specifically, we propose an approach that explicitly models the existence of two interacting groups of heterogeneous agents, to simultaneously learn and optimize the overall system performance. Namely, a group of autonomous end-users that aim to perform network selection and a group of autonomous RANs that seek to solve the problem of resource allocation. Note that this contrasts the previous studies cited in the paper as those focus on optimizing the task of one group only, while considering the other as static respondents or assume a centralized control that optimizes the performance of the two groups. Our paper is the first to explicitly model the adaptive and decentralized behavior of these two groups, where each one is learning simultaneously and independently from the others. Moreover, the proposed framework is supported by a practical and essential application in smart health systems, which are perfect candidates to leverage the advantages brought by the framework. The entities of these systems dynamically change, e.g., the patients' health state, the wireless channel characteristics, and dynamics, the battery level of the mobile device...etc.
Our main contributions can be summarized as follows:
\begin{enumerate}
\item We formulate a multi-objective optimization problem that describes the whole system and aims at obtaining, for each user/patient edge node (PEN): (i) the optimal data compression ratio, (ii) the selected Radio Access Networks (RANs) for data transmission, and (iii) the allocated bandwidth at each selected RAN(s). 

\item The problem is reformulated in terms of Multi-Agent theory, where the agents for each interacting group in the system (i.e., RANs and PENs) will be defined along with their observations, actions and their reward functions.

\item We propose a distributed DMARL algorithm, namely, Team-Based Multi-Agent Deep Deterministic Policy Gradient (TB-MADDPG), which handles the heterogeneity of the agents in the system, and looks for the optimal joint policy that maximizes the rewards for all agents. 


\end{enumerate}	

The rest of the paper is organized as follows. Section \ref{sec:sm} introduces the system model and the main Performance metrics considered in this study. Section \ref{sec:formulation} presents the formulated optimization problem, along with the re-formulation in terms of multi-agent theory. In Section \ref{sec:solution_approaches}, we present the proposed approach to solve the reformulated problem and learn the optimal policy. Section \ref{sec:results} presents the performance evaluation of our approach and the comparisons against the state-of-the-art techniques, while Section \ref{cmplx_analysis} presents the complexity analysis of the proposed method before concluding in Section \ref{sec:conclusion}.

\section{System model and Performance metrics}
\label{sec:sm}
This section introduces the system model under study. Then, it presents the main application and network requirements that will be considered in the proposed framework.

\begin{figure}
	\centering
	\includegraphics[scale=.87]{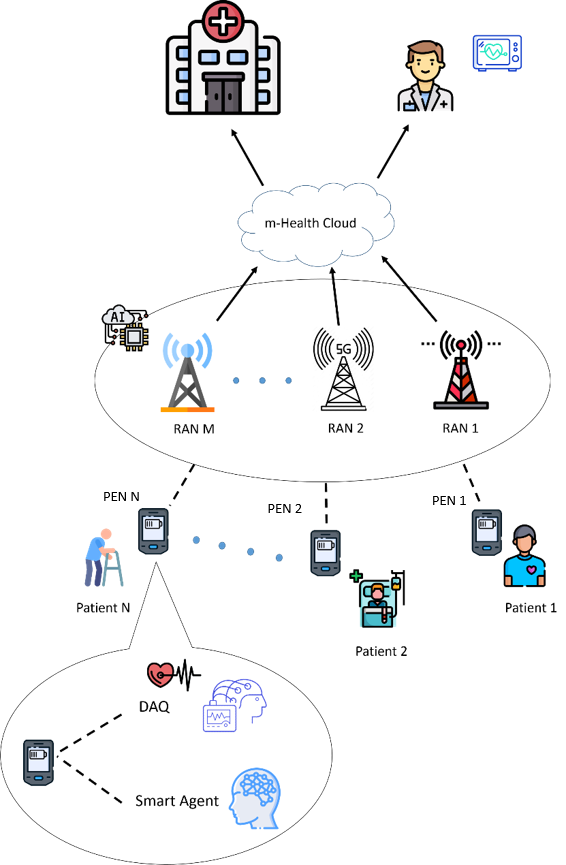}
	\caption{\footnotesize System model under study}
	\label{fig:sys_model_fig}
\end{figure}

\subsection{System Model}
In this work, we consider a mobile-health (m-health) network with ultra-dense heterogeneous network architecture, where multiple end-users can access multiple RANs as in Fig \ref{fig:sys_model_fig}. In the considered scenario, a combination of sensor nodes attached to the patients is used for monitoring the patient’s health state (e.g., implantable or wearable sensors that measure various biosignals and vital signs). The acquired data from such sensors are forwarded to a patient edge node (PEN) that represents the data processing unit between the data sources and the RANs.  Specifically, the PEN collects the medical data from various sources and applies an adaptive compression scheme to control the data size, while considering the high-level application's requirements. Then, the PEN can forward the collected data to the m-health cloud (MHC) via the available RANs. Each RAN has different characteristics, such as data rate, energy consumption, monetary cost (i.e., requested payment for using network services), and transmission delay. Moreover, due to patient mobility, the level of quality of service offered by the available RANs may vary over time.

The PENs are considered to be battery-operated devices, where a PEN can be any mobile node (e.g., smartphone). However, it can collect an immense amount of data due to the continuous monitoring of diverse patient's conditions. Hence, it is important to transmit the collected data efficiently through the RANs without draining the PEN's battery, such that the PEN can be used for a long time without charging or replacement.

Lastly, we consider that the patient's conditions might change at any point in time in order to simulate a real-life scenario. More specifically, the patients in our system model can have abnormal conditions (such as seizure) for a period of time, and then their conditions can go back to normal after a certain time period. Such abnormal conditions influence how the vital signs should be processed locally at the edge on one side, and what RANs will be selected to address the emergency case with ultra-low latency on the other side.

\subsection{Performance Metrics}

In the following, we assume on a time period $T$, each PEN $i$ ($i=1,...N$) has to transfer $B_i$ bits of data towards the MHC, through $M$ RANs. Moreover, each RAN $j$ ($j=1,...M$) has to allocate resources in terms of bandwidth to $N$ PENs.

\textbf{Adaptive Compression:}   
The adaptive data compression at the PEN, is implemented using the discrete wavelet transform (DWT) compression scheme \cite{DWT_eeg}.  Hence, the generated data length, after compression, at PEN $i$ is expressed as follows:
\begin{equation}
b_i = B_i(1- \kappa_i)
\end{equation}
where $B_i$ being the length of the raw data sent by PEN $i$ before compression, and $\kappa_i$ is the compression ratio at that PEN. However, using lossy data compression, prior to the transmission, comes at the cost of introducing data distortion at the receiver side. In the proposed framework, we adopt the compression scheme in \cite{Real_time_awad} for electroencephalogram (EEG) data compression, as an example of intensive medical data compression. Nevertheless, without loss of generality, the proposed framework can be easily extended to consider different compression schemes, medical or non-medical data (e.g, video) \cite{ABDELLATIF202053}.  Using the obtained results in \cite{Real_time_awad}, the introduced distortion can be expressed as: 
\begin{equation}\label{eqn:eqndistortion}
D_i =\dfrac{c_1. \exp(1-\kappa_i) + c_2. (1-\kappa_i)^{-c_3} + c_4 .F^{-c_5}-c_6}{100},
\end{equation}
where $F$ is the wavelet filter length of the adopted DWT compression scheme, and $c_1-c_6$ are the estimated parameters by the statistics of the EEG compression model.

\textbf{Energy Consumption:} First, the available data rate from RAN $j$ to PEN $i$ is given by: 
\begin{equation}
r_{ij} = W_{ij} \log_2 \big( 1 + \frac{P_t \, g_{ij}}{N_0 W_{ij}} \big)
\end{equation}
where $P_t$ is the PEN transmission power, $W_{ij}$ is the allocated bandwidth, and is given by $W_{ij} = \theta_{ij}.W_j$, with $\theta_{ij}$ being the fraction of the RAN bandwidth $W_j$ \cite{5982690}. In (3), $N_0$ is the noise spectral density, while the channel gain $g_{ij}$ is defined as:
\begin{equation}
g_{ij} = K ~.~ \sigma ~. ~|h_{ij}|^2 
\end{equation}
where K = -1.5/(log(5BER)), $\sigma$ is the path loss attenuation, and $|h_{ij}|$ is the fading channel magnitude for PEN $i$ over RAN $j$.
Then, the estimated energy consumption at PEN $i$ to send $b_i$ bits over RAN $j$, as defined in \cite{awad2014interference}, is given by:
\begin{equation}
E_{ij} = \psi_j \, . \, \bigg ( \frac{b_i N_0 W_{ij}}{r_{ij}g_{ij}} (2^{\frac{r_{ij}}{W_{ij}}}-1) \bigg) + c_j,
\label{eqn:energy}
\end{equation}
where $\psi_j$ and $c_j$ are specific parameters that differ for each network interface \cite{mahmud2004measurement}. 

\textbf{Latency:} The expected latency to send $b_i$ bits from PEN $i$ through RAN $j$ can be defined as:
\begin{equation}
L_{ij} = \frac{b_i}{r_{ij}} + \xi_j
\end{equation}
where $\xi_j$ is the access channel delay over RAN $j$. More specifically, the expected latency represents the end-to-end delay when using a given technology \cite{wang2011cross}.
 Nonetheless, the data rate equation in (3) is generic and interference-free, and any other data rate model with interference can be simply integrated along with the data transmission energy consumption model.

\textbf{Monetary Cost:} the cost resulting from using RAN $j$ by PEN $i$ to send $b_i$ bits is expressed in Euro's and can be defined as:
\begin{equation}
C_{ij} = b_i\varepsilon_j
\end{equation}
where $\varepsilon_j$ is the monetary cost per sent bit over RAN $j$. This monetary cost can be acquired through the use of, e.g., the IEEE 802.21 standard \cite{ieee2009ieee}, which allows a user device to gather information about the available wireless networks. Such value can also be stored in the PENs in advance and updated if there are any changes in pricing.

\section{Problem formulation}
\label{sec:formulation}

In this section, firstly, we formulate our problem for optimal Network selection and resource allocation as a multi-objective
optimization problem under certain network constraints. Secondly, we motivate and introduce a necessary reformulation as a MARL framework by expressing the problem as a discrete time-variant system.

\subsection{Multi-objective optimization problem formulation}
The goal of the optimization problem is to minimize the transmission energy consumption, along with the data delivery latency and monetary cost, while meeting the medical data QoS requirements in terms of distortion. Therefore, we define an objective function that is a weighted sum aggregate of the aforementioned objectives. 

Given $N$ PENs with $M$ available RANs, the objective of our optimization problem is to minimize the PENs' transmission energy consumption $E_{ij}$ , monetary  cost $C_{ij}$ , latency $L_{ij}$ , and distortion $D_i$.  
This can be achieved through:  (i) assigning the PENs to the optimal RAN(s), (ii)  obtaining the optimal bandwidth allocation over different RANs, and (iii) setting the PENs' optimal compression ratio. Thus, the proposed optimization problem is formulated as: 

\begin{equation}\label{eqn:objective}
\textbf{P1}:~~\min_{\boldsymbol{\theta,}\mathbf{ P,\kappa}} \,\,\, \mathbb{E} \bigg [ \sum_i \sum_j P_{ij} U_{ij} + \delta_i D_i \bigg ]
\vspace{-0.1in}
\end{equation}
\begin{eqnarray*}
s.t~~~~~~~~~~~~~~~~~~~~~~~~\;\;\;\;
\vspace{-0.1in}
\end{eqnarray*}
\begin{equation}
\sum_i \theta_{ij}= 1 , \,\;\; \forall j \in M\\
\end{equation}
\begin{equation}
\sum_j P_{ij}= 1 , \,\;\; \forall i\in N\\
\end{equation}
\begin{equation}\label{eqn:Delay_const}
\frac{b_i P_{ij}}{r_{ij}} \leq T_{ij} , \,\;\; \forall i \in N, \forall j\in M\\
\end{equation}
\begin{equation}
0 \leq \theta_{ij}, P_{ij} \leq 1 , \,\;\; \forall i \in N, \forall j\in M 
\end{equation}
\begin{equation}
0 \leq \kappa_i < 1,~~ \forall i \in N~~~~
\end{equation}

where $U_{ij} = \alpha_i E_{ij} + \beta_i C_{ij} +  \lambda_i L_{ij}$ is the utility function of PEN $i$ over RAN $j$. The weighting coefficients $\alpha$, $\beta$, $\lambda$, and $\delta$ represent the
relative significance of the four metrics in the problem;
where $ \alpha_i+\beta_i+\lambda_i+\delta_i = 1 $. Moreover, we denote the patient status as $\zeta_i$, where:
\begin{equation}\label{eqn:seizure_prop}
\textbf{P}(\zeta_i = 1) = \upsilon_i, ~~ \forall i \in N    
\end{equation}
where the equation conveys that a patient $i$ can have a seizure at any point in time with a probability of $\upsilon_i$. The weighting coefficients $\alpha$, $\beta$, $\lambda$, and $\delta$ are functions of $\zeta_i$, where they can have different values according to the patient status, which allows us to optimize the RAN(s) selection and PENs' parameters based on the patient status. Since the weighting coefficients are a function of the random variable $\zeta_i$, and hence, the objective function is stochastic, we optimize the expected value of the objective function. Moreover, It is worth noting that the metrics are normalized between 0 and 1 in order to sum the unitless values in the objective function (\ref{eqn:objective}).

In (\ref{eqn:objective}), we consider a network utilization indicator $P_{ij}$ that represents
the fraction of data that should be transmitted through RAN $j$ by PEN $i$. We assume that the PENs have all information to compute the energy consumption and cost.  
The constraint in (9) ensures the full utilisation of RANs' bandwidth by the PENs, while the constraint in (10) ensures that all the data that each PEN has to transfer to the MHC is actually sent through the RANs. The network capacity constraint is represented by (\ref{eqn:Delay_const}), where $T_{ij}$ is the maximum fraction of the time period $T$ that can be used by PEN $i$ over RAN $j$ (i.e., resource share). $T_{ij}$ depends on the number of PENs accessing the RAN, and we assume that it is notified by the RANs. 

The decision variables in this problems are the $\theta_{ij}$'s , $P_{ij}$'s and the $\kappa_i$, i.e., each RAN needs to determine the allocated bandwidth for each PEN, and each PEN needs to determine its compression ratio and the amount of data it should transfer through  different RANs. However, the problem $\textbf{P1}$ in (\ref{eqn:objective}) in its current form is not convex \cite{boyd2004convex}, since the second derivatives matrix of the  utility function, i.e., the Hessian matrix, is not positive semi-definite.  
Moreover, an approach like transforming the problem into a Geometric Program (GP) would not work in this case due to the existence of the non-linearity of the distortion in the objective. 
However, the problem can be solved in a distributed manner where it can be divided into sub-problems and the variables can be separated, which the authors in \cite{ONSRA} have done. Nevertheless, the obtained solution does not take into accounts the time variance in the system, and solving the problem $\textbf{P1}$ by classical optimization approaches as in the previous work is computationally expensive, since the system parameters change dynamically with time (e.g., channel gain of each RAN, patients' conditions, PENs battery status...etc.). Also, with changing the environment, the objective function has to be re-optimized again considering the new changes. Thus, in what follows, we opt to reformulating the problem, leveraging multi-agent systems theory, to be solved using deep MARL (DMARL)-based approach. 

\subsection{Multi-Agent Reinforcement Learning Formulation}
As mentioned earlier, we want to solve the problem in an online manner, adapting to the changes in real-time. However, since the problem is non-convex, solving it repeatedly at every variation of the system parameters is inefficient.
 Hence, we opt for learning-based optimization such as reinforcement learning (RL) to achieve this adaptation. To solve the problem through RL, a Partially-Observable Markov Decision Process (POMDPs) representation of the system has to be designed. In fact, due to the existence of more than one independent PEN in the system along with multiple RANs calls for the need of the multi-agent extension of POMDPs.
In our new formulation, we consider multiple heterogeneous agents interacting  simultaneously with a partially observable environment $V$ in discrete time steps. The formulation is to be described by a partially observable Markov game (POMG)\cite{POMG}, which is a multi-agent extension for POMDPs. POMGs are usually represented by the tuple $(\mathcal{N},\mathcal{S},\mathcal{A},\mathcal{O},\mathcal{T},\mathcal{R}, \gamma)$, where $\mathcal{N}$ is the set of all agents, $s_t \in \mathcal{S}$ is a single possible configuration of all the agents at time $t$, $a_t \in \mathcal{A}$ is a possible action for the agents where $\mathcal{A} = \mathcal{A}_1 \times \mathcal{A}_2 \times ....\times \mathcal{A}_N$, $o_t \in \mathcal{O}$ is a possible observation of the agents where $\mathcal{O} = \mathcal{O}_1 \times \mathcal{O}_2 \times .... \times \mathcal{O}_N$, $\mathcal{T}$ is the state transition probability $\mathcal{T}: \mathcal{O} \times \mathcal{A} \mapsto \mathcal{O}$, $\mathcal{R}$ is the set of rewards for all the agents $r: \mathcal{O} \times \mathcal{A} \mapsto \mathbb{R}$, and $\gamma$ is the reward discount factor. In what follows, we define each set in the tuple in the context of our network selection and resource allocation problem. 
\hfill\\

\subsubsection{System Agents and Environment} \hfill\\ \vspace{-.1in}

In this work, we consider two types of agents, namely, the PEN agents and the RAN agents. The PEN agents are deployed at the PENs and can control the data transmission and compression, while the RAN agents are deployed in the RAN controller in order to optimize the RANs' bandwidth allocation to the PENs. Each agent $i$ receives from the environment an observation, and take actions according to the joint policy $\boldsymbol{\pi}: \mathcal{O} \mapsto \mathcal{A}$, where $\boldsymbol{\pi}: \pi_1 \times \pi_2 \times.... \times \pi_N$, and it transits the agent to its next state according to the state transition probability $\mathcal{T}$, and the agent receives an immediate reward $r_i$. The environment $V$ will be episodic, i.e., it has finite horizon, and it terminates when all the PENs run out of power. During an episode, the  total cumulative discounted reward for an agent $i$ is denoted by $R_i = \sum_{t=0}^{t'} \gamma^t r_i^t $, where $\gamma \in [0,1)$ and $t'$ is the episode time horizon. The state-action value function of a joint policy $\boldsymbol{\pi}$ for an agent $i$ is expressed as $Q_i^{\boldsymbol{\pi}}(s,a) = \mathbb{E} \left [ R^t_i| s^t=s, a^t=a\right ]$. The goal of the agents is to find an optimal joint policy  $\boldsymbol{\pi}^*$ that yields the optimal Q-function $Q_i^{\boldsymbol{\pi}^*}(s,a)=\max_{\boldsymbol{\pi}} Q_i(s,a)$ for each agent through direct interactions with the environment, without having an explicit pre-knowledge about it (the transition probability $\mathcal{T}$).
\hfill\\
\subsubsection{Agents Observations and Actions} \hfill\\\vspace{-.1in}

In our system, we consider two types of agents, namely, the RAN agent type and the PEN agent type. Each type of agents has different observations and actions, and each agent from each type has his own observations and actions. We assume that agents from the same type are independent from each other, and they do not share any observation or any kind of information. In fact, this is the case in real-world scenarios, where PENs as actors do not share any information with each other in order to preserve their privacy, and each RAN does not have any information about other RANs and how much data have been flowed through them. However, agents from different type have to share some information between each other, including whether PENs are willing to send any data to a RAN or not, and how much bandwidth is each RAN allocating to each PEN. We denote the RAN agent $j$ observation and action at time $t$ by $o_{\rho_j}^t$ and $a_{\rho_j}^t$ respectively, and the PEN agent $i$ observation and action at time $t$ by $o_{\varphi_i}^t$ and $a_{\varphi_i}^t$ respectively. The RAN agent $j$ observation at time $t$ is the status of the PENs if they have sent any data or not to RAN $j$, and given by:
\begin{equation}
o_{\rho_j}^t = \{ n^{t}_{ij} \} ,\,\,\,\, \forall i \in N
\end{equation}
where,
\begin{equation}
n^t_{ij}=\left\{\begin{matrix}
 1& P^{t}_{ij}>0\\ 
 0& P^{t}_{ij}=0,
\end{matrix}\right.
\end{equation}
whereas the RAN agent action is how much bandwidth it allocates to each PEN, or the fraction of the RAN bandwidth that each PEN is getting at time $t$, and given by: 
\begin{equation}
a_{\rho_j}^t = \{ \theta^t_{ij}\} ,\,\,\,\, \forall i \in N
\end{equation}

As for a PEN agent $i$, its observation at time $t$ consists of the PEN energy consumption, latency, monetary cost and the channel state for each RAN that the PEN has sent any data through. Moreover, the PEN agent observes the patient seizure status if it is active or not, along with the medical data distortion and the PEN battery level. The full PEN agent observation can be expressed as the following:
\begin{equation}
o_{\varphi_i}^t = \{ \bar{E}^{t}_{ij}\,,\, \bar{C}^{t}_{ij}\,,\, \bar{L}^{t}_{ij} ,\, \bar{D}^t_i\,,\, \zeta^t_i\,,\, \bar{\Gamma}^t_i\}, \,\,\,\, \forall j \in M\
\end{equation}
where $\bar{E}^{t}_{ij}, \bar{C}^{t}_{ij}, \bar{L}^{t}_{ij}, \bar{D}^t_i$ and $\bar{\Gamma}^t_i$ are the normalized energy consumption, monetary cost, latency, distortion and battery level respectively, by max normalization (i.e. they fall in the range of $[0-1]$). Moreover, $\bar{E}^{t}_{ij}$, $\bar{C}^{t}_{ij}$ and $\bar{L}^{t}_{ij}$ at time $t$ can be evaluated by plugging $P^{t-1}_{ij}$ and $\theta^{t-1}_{ij}$ in (5), (6) and (7) respectively, whereas $\bar{D}^t_i$ by plugging $\kappa_i^{t-1}$ in (2), and finally, $\bar{\Gamma}^t_i$ is evaluated as follows:
\begin{equation}
\bar{\Gamma}^t_i = \frac{\Gamma^{t-1}_i-\sum_j E^t_{ij}}{\Gamma^0_i}
\end{equation}
It is worth pointing out that the transitions of $\bar{E}^{t}_{ij}$, $\bar{L}^{t}_{ij}$ and $\bar{\Gamma}^{t}_{i}$ are stochastic, due to the fact that these observations mainly depend on the channel gain estimation, which introduces the randomness in the system. Also, the patient status is not deterministic, as the patient is prone to abnormal conditions (i.e., seizures) at any point in time. \hfill\\
Lastly, the PEN agent $i$ actions at time $t$ are the amount of data that the PEN is sending to each RAN, along with the compression ratio, and can be expressed as:
\begin{equation}
a_{\varphi_i}^t = \{ P^t_{ij},  \kappa_i \},\,\,\,\,   \forall j \in M
\end{equation}
\subsubsection{Agents Reward functions}
\hfill\\\vspace{-.1in}

The reward functions have to be designed such that they describe the original optimization problem along with maximizing the PENs battery lifetime. The PEN agents aim is to minimize the energy consumption in order to guarantee a longer battery lifetime, while jointly minimizing the cost, latency and distortion. Therefore, the PEN agent reward function is defined as the following:
\begin{equation}\label{eqn:reward_fun}
\small
r^t_{\varphi_i}=\left\{\begin{matrix}
(1-\sum_j P^t_{ij}\bar{U}^t_{ij})+\delta_i(1-\bar{D}^t_i)+ \bar{\Gamma}^t_i & \small \zeta^t_i = 0\\ 
\lambda_i(1-\sum_j \bar{L}^t_{ij})+\delta_i(0.1-\bar{D}^t_i)+ \bar{\Gamma}^t_i & \small \zeta^t_i = 1\\ 
 -1& ~\text{violated constraints}
\end{matrix}\right.
\end{equation}
\begin{figure*}[!t]
	\centering
	\includegraphics[scale=.27]{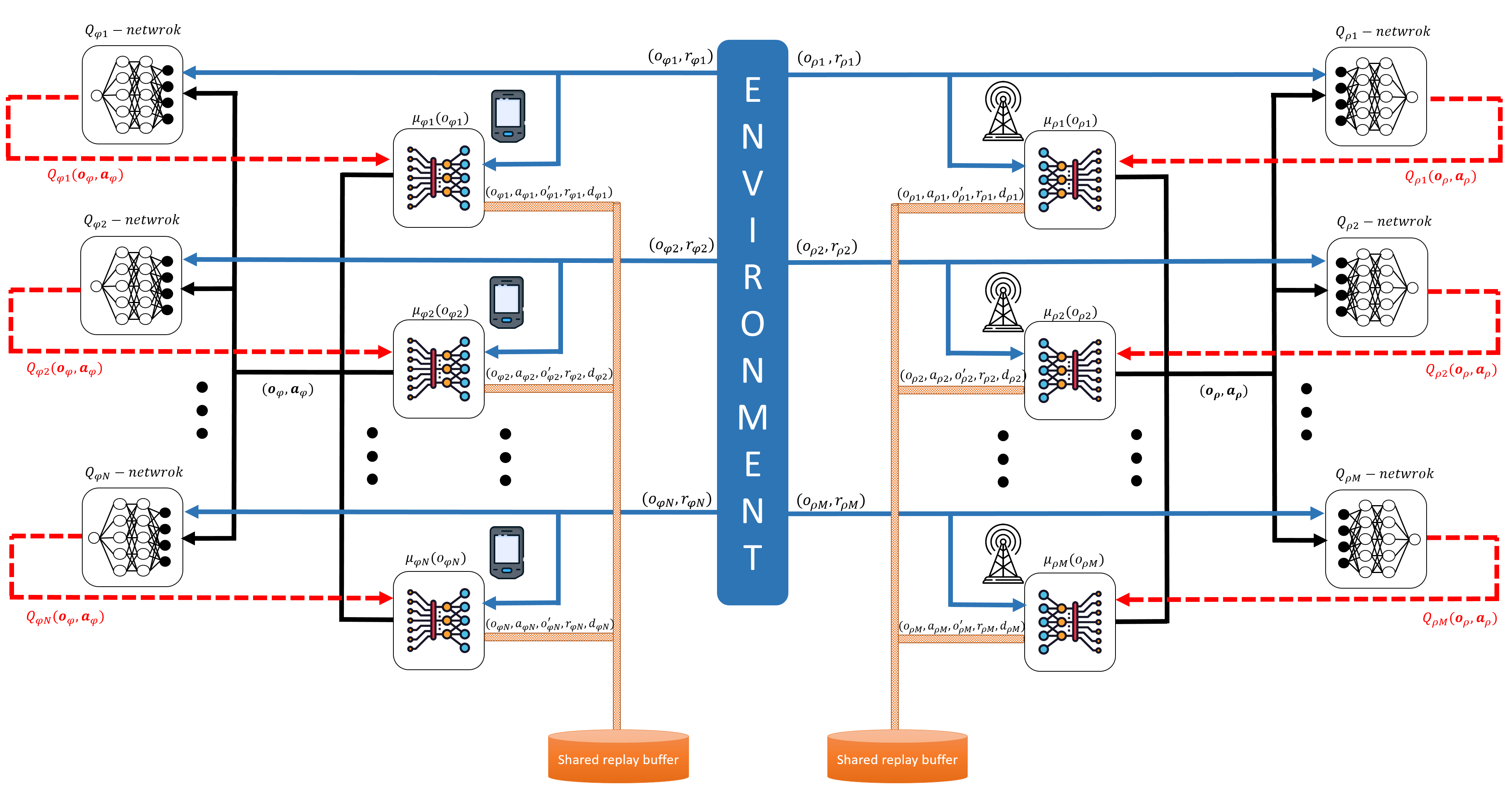}
	\caption{\footnotesize TB-MADDPG training architecture}
	\label{fig:training_arch}
\end{figure*}

\begin{figure}
	\centering
	\includegraphics[scale=.25]{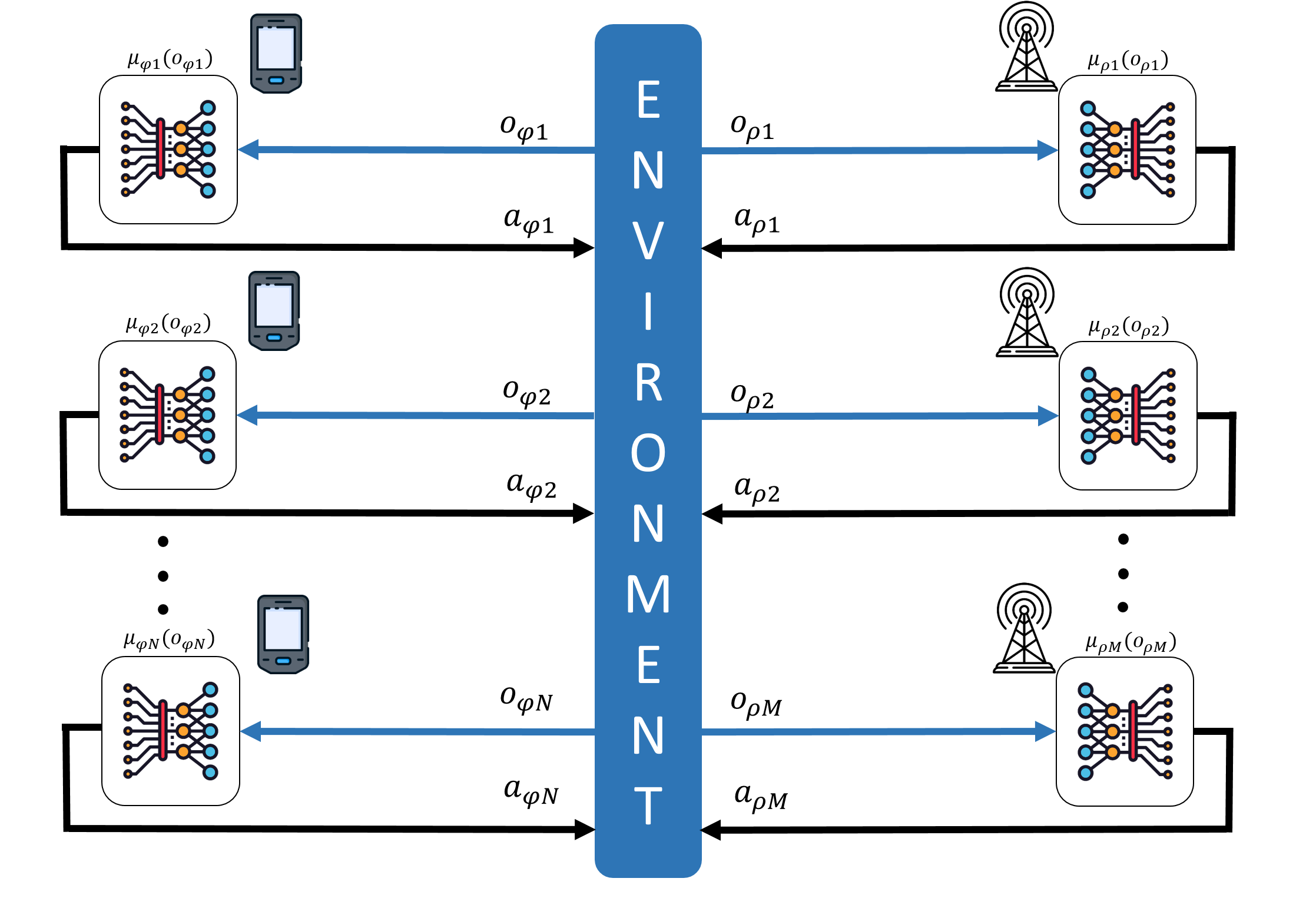}
	\caption{\footnotesize TB-MADDPG execution architecture}
	\label{fig:execution_arch}
\end{figure}

where $\bar{U}^t_{ij}$ is the normalized utility function (i.e., the sum of the normalized energy consumption, monetary cost and latency). The first and the second term represents the objective function (\ref{eqn:objective}), while the third term addressed the battery energy level, and it forces the agent to minimize the energy consumption in order to maintain a slow decrease in the reward over time, rather than a sharp decrease. However, when the patient has a seizure, the data has to be delivered with the minimum distortion and delay possible. Hence, the reward will disregard the cost factor, and care less about the energy consumption, and give more significance to the latency and distortion terms. Lastly, if the agent violates any constraints, it will receive a penalty of -1.

The RAN agents' main goal is to maximize the connected users QoE, which can be done by maximizing their rewards. Hence, the reward for the RAN agents is defined as the following:
\begin{equation}
r^t_{\rho_j}=\left\{\begin{matrix}
\frac{\sum_i n^t_{ij}r^t_{\varphi_i}}{\sum_i n^t_{ij}} & (9),(11)~\text{hold}\\ 
 -1& ~\text{otherwise}
\end{matrix}\right.
\end{equation}
where $n^t_{ij}$ is the status of PEN $i$ if it has sent any data to RAN $j$ at time $t$ and $r^t_{\varphi_i}$ is the reward of the PEN agent $i$. The aforementioned reward function represents the average rewards of the connected PENs to the RAN $j$. The agents at each RAN will try to maximize the rewards of its connected PENs by allocating the optimal bandwidth for each one, in order to guarantee the best QoE for the connected users.

\section{Team-Based MADDPG Solution For Network Selection and Resource Allocation}\label{sec:solution_approaches}
MADDPG \cite{lowe2020multiagent} is the multi-agent extension of the DDPG \cite{lillicrap2019continuous} algorithm, where each agent learns a policy based on its local observation only, and each agent has his own Q-function, which works as an indicator of how good is the agent policy. The MADDPG algorithm adopts the centralized training with decentralized execution framework. In the training phase, which is depicted in in Fig. \ref{fig:training_arch}, each agent utilizes a Q-function (critic), and a policy function (actor), where the critics have access to the joint observations and actions (hence the name ``centralized"), and guides the training of the actors. In the testing (execution) phase, shown in Fig. \ref{fig:execution_arch}, the critic is discarded (i.e., the global observation/actions are not needed any more) and only the actor, which depends only on local observations, is utilized. The centralized training allows the policies to make use of additional information to ease and stabilize training, so long as this information is not used at execution time. This adopted framework in MADDPG leads to learned joint optimal policies that only use local information (i.e., their own observation only) at execution time. Moreover, MADDPG is not only limited to a single type of interaction between the agents, but can use any type of interaction including cooperative, competitive, mixed or neither of these types.

However, the MADDPG assumes the agents are homogeneous (i.e., they have the same structure of observation and actions), which is not the case in our environment, where we have two types of agents, with different structure of observations and actions. Moreover, having heterogeneous agents in the environment makes the environment even less stationary, which hardens the agents-learning process from the environment interactions. Hence, we propose the Team-Based MADDPG (TB-MADDPG), which is essentially a dual MADDPG frameworks interacting with each other, and groups each type of agents in a team, and each team has a different joint policy, and therefore, a different Q-function to assess each joint policy.
The first MADDPG has the PENs as its agents, and considers the other team as part of the environment. Similarly, the second MADDPG has the RANs as its agents, and considers the PENs as part of its environment.
Furthermore, each team make use of a shared replay buffer that stores all the agents' experiences. An agent experience can be represented by the tuple $(o,a,o',r,d)$, where $d$ represents the done flag (i.e., when the agent stops interacting with the environment). The agents from the same team are not allowed to observe other agents experiences in the shared replay buffer. However, the critic networks make full use of that buffer in order to assess the joint actions given a joint observation, and hence, assessing the joint policy. Such modification makes the environment more stationary, and make the learning in each team as a normal homogeneous MADDPG training.

As mentioned before, there will be two teams in our solution design, namely, the RANs and the PENs teams. We denote the joint observation and joint action for the RAN team by $\boldsymbol{o_\rho}$ and $\boldsymbol{a_\rho}$ respectively, where a single action for an agent ${a_{\rho_j}} = \mu_{\rho_j}(o_{\rho_j} ; \vartheta_{\rho_j})$, with $\mu_{\rho_j}(.)$ denoting the output of the RAN actor network $j$, and $\vartheta_{\rho_j}$ is the set of parameters of the actor network.
Similarly for the PENs team, we denote the joint observation and joint action by $\boldsymbol{o_\varphi}$ and $\boldsymbol{a_\varphi}$ respectively, and single action for the PEN agent $i$ is  ${a_{\varphi_i}} = \mu_{\varphi_i}(o_{\varphi_i} ; \vartheta_{\varphi_i})$.
Also, the joint policy for the RANs team and the PENs team are denoted by $\boldsymbol{\pi_\rho}$ and $\boldsymbol{\pi_\varphi}$. Finally, a single Q-function in the RANs team and the PENs team is denoted by $Q_{\rho_j}(\boldsymbol{o_\rho}, \boldsymbol{a_\rho}; \phi_{\rho_j})$ and $Q_{\varphi_i}(\boldsymbol{o_\varphi}, \boldsymbol{a_\varphi}; \phi_{\varphi_i})$, where $\phi_{\rho_j}$ and $\phi_{\varphi_i}$ are the set of parameters for the critic network $j$ and $i$ in the RANs and the PENs team respectively.

\begin{algorithm}[h]
\small
\SetAlgoLined
Initialize actors and critics networks parameters\\
Initialize a random process $\mathcal{N}$ for action exploration\\
 \For{episode e=1:Episodes}{
     Receive initial states for both teams $\boldsymbol{o}_\rho, \boldsymbol{o}_\varphi$\\
    \For{time step t=1:Steps}{
        \textbf{At the PENs side:}\\
        \For{$\text{PEN agent}$ i=1:N}{
            Select and execute action $a_{\varphi_j} = \mu_{\varphi_j}(o_{\varphi_j}) + \mathcal{N}$ following policy $\boldsymbol{\pi}_\varphi$ \\
        }
        Observe PENs next states $\boldsymbol{o}'_\varphi$ and rewards $\boldsymbol{r}_\varphi$\\
        Store $(\boldsymbol{o}_\varphi, \boldsymbol{a}_\varphi, \boldsymbol{o}'_\varphi, \boldsymbol{r}_\varphi)$ in PENs team buffer $\mathcal{D}_\varphi$\\
        \textbf{At the RANs side:}\\
        \For{$\text{RAN agent}$ j=1:M}{
            Select and execute action $a_{\rho_j} = \mu_{\rho_j}(o_{\rho_j}) + \mathcal{N}$ following policy $\boldsymbol{\pi}_\rho$ \\
        }
        Observe RANs next states $\boldsymbol{o}'_\rho$ and rewards $\boldsymbol{r}_\rho$  \\
        Store $(\boldsymbol{o}_\rho, \boldsymbol{a}_\rho, \boldsymbol{o}'_\rho, \boldsymbol{r}_\rho)$ in RANs team buffer $\mathcal{D}_\rho$\\
        \textbf{At both sides:}\\
        \If{time to train}{
        Sample a mini-batch $\mathcal{B}_x$ from each $\mathcal{D}_x$\\
        \For{$\text{each agent}$}{
            Compute targets $y_{x}(\boldsymbol{o'}_{x},r_{x},d_{x})$ according to (\ref{eqn:TD})\\
            Compute loss function $\mathcal{L}(\phi_{x}, \mathcal{B}_x)$ according to (\ref{eqn:loss})\\
            Update critic parameters: \\
            $\phi_{x} \leftarrow \phi_{x} -\eta_{\phi_{x}} \nabla_{\phi_{x}}\mathcal{L}(\phi_{x},\mathcal{B}_{x})$\\
            Update actor parameters: \\
            $\vartheta_x \leftarrow \vartheta_x +\eta_{\vartheta_x} \nabla_{\vartheta_x}\underset{o_x\sim \mathcal{D}_x}{\textrm{E}}  \bigg [ Q_x\big(o_x,\mu_x(o_x;\vartheta_x) ; \phi_x\big)  \bigg]$\\
            Update target networks parameters according to (\ref{eqn:update_phi}), (\ref{eqn:update_vartheta})
        }
        }
    }
}
\caption{\small Team Based-Multi Agent DDPG}
\label{algorithm}
\end{algorithm}
Learning in TB-MADDPG consists of Q-learning for the critics, and policy learning for the actors. The concept of Q-learning is that if the optimal Q-function is known, then the optimal action to take is the one which maximizes the Q-function. Since each critic parametrize the Q-function, the goal of the critic is to be as close as possible to the optimal Q-function. Therefore, for each critic, we can set up an indicator of how close the critic to the optimal Q-function is, which is the Mean-Squared Bellman Error (MSBE), and defined as the following:
\begin{equation}\label{eqn:loss}
\mathcal{L}(\phi_x, \mathcal{B}_x)  =  \underset{\mathcal{B}_x\sim \mathcal{D}_x}{\textrm{E}} \Bigg [   \Big ( Q_x(\boldsymbol{o}_x,\boldsymbol{a}_x; \phi_x) - y_x(\boldsymbol{o'}_x,r_x,d_x)    \Big )^2 \Bigg ]
\end{equation}
where $x\in \{\rho, \varphi\}$, $\mathcal{D}$ is the shared replay buffer in the team, and $\mathcal{B}$ is a sampled batch of joint experiences from the buffer. The term $y_x(.)$ is known as the Temporal Difference (TD)-target, and is given by:
\begin{equation}\label{eqn:TD}
y_x(\boldsymbol{o'}_x,r_x,d_x) =  \Big(~r_x+ \gamma(1-d_x) Q_x\big(\boldsymbol{o'}_x,\boldsymbol{a'}_x; \phi_x\big) \Big)
\end{equation}
where $\boldsymbol{a'}$ is the joint action from the actors given the next state $\boldsymbol{o'}$.
The aim of each critic is to minimize the MSBE, which can be done by gradient descent on its parameters $\phi$:
\begin{equation}
\phi_x \leftarrow \phi_x -\eta_{\phi_x} \nabla_{\phi_x}\mathcal{L}(\phi_x,\mathcal{D}_x)
\end{equation}
where $\eta_{\phi}$ is the learning rate of the critic network.

Each actor in TB-MADDPG learns independently, relying on its local observations and its critic only, where the actor tries to learn an optimal policy that maximizes the Q-function. Hence, the objective of each actor from both teams can be expressed as: 
\begin{equation}
\max_{\vartheta_x}  \underset{o_x\sim \mathcal{D}_x}{\textrm{E}}  \bigg [ Q_x\big(o_x,\mu_x(o_x;\vartheta_x) ; \phi_x\big)  \bigg]
\end{equation}
which can be maximized by simple performing gradient ascent on the actor parameters $\varphi$ as follows:
\begin{equation}
\vartheta_x \leftarrow \vartheta_x +\eta_{\vartheta_x} \nabla_{\vartheta_x}\underset{o_x\sim \mathcal{D}_x}{\textrm{E}}  \bigg [ Q_x\big(o_x,\mu_x(o_x;\vartheta_x) ; \phi_x\big)  \bigg]
\end{equation}
where $\eta_{\vartheta}$ is the learning rate of the actor network. Moreover, to stabilize the learning process \cite{experience_replay}, each actor and critic network have a time-delayed copy of itself, and are called the target networks. The parameters of the target networks are updated as follows: 
\begin{equation}\label{eqn:update_phi}
\phi_{targ} \leftarrow (1-\epsilon)\phi_{targ} + \epsilon\phi
\end{equation}
\begin{equation}\label{eqn:update_vartheta}
\vartheta_{targ} \leftarrow (1-\epsilon)\vartheta_{targ} + \epsilon\vartheta
\end{equation}
where $\epsilon$ is the soft update parameters, and $\epsilon << 1$. The TB-MADDPG algorithm is summarized in Algorithm \ref{algorithm}. 

Lastly, it is worth recalling that the TB-MADDPG algorithm is done with centralized training (Fig \ref{fig:training_arch}) and decentralized execution (Fig \ref{fig:execution_arch}). In other words, during training, the agents' critics can make use of extra knowledge about other agents' interaction with the environment in order to help the actors to learn the optimal joint policy. However, at inference time, the actors only receive their local observation and do not share or exchange any private information with other agents. The online training can be held in a trusted virtualized environment, where the replay buffers of each team can be stored and can receive the experiences from the agents, and deploy the models on the PENs and the RANs every once in a while.

\section{Simulation results}\label{sec:results}
In this section, we first present the environmental setup. Second, we evaluate the agents' behavior in terms convergence and the learned policy. Lastly, we compare our proposed approach to existing state-of-art techniques for network selection and resource allocation.

\subsection{Environmental Setup}
\begin{table}[t]
\caption{Simulation parameters}
 \label{table:parameters}
\centering
\begin{tabular}{ll}
\hline
Parameter & Value \\
\hline
Number of RANs $M$  &       3\\ \hline
Number of PENs $N$  &       5\\ \hline
Probability of seizure $\upsilon$  &       0.1\\ \hline
Data length $B$  &       1 Mb\\ \hline 
Distortion parameters $c1-c6$  &       as in \cite{awad2014real_ccc}\\ \hline 
RAN types   &      [5G, 4G, 3G] \\ \hline 
Data rates $r_1, r_2, r_3$   &      [40, 25, 15] Mbps \\ \hline
Monetary costs $\varepsilon_1, \varepsilon_2, \varepsilon_3$    &    [6, 3, 0.1]$*10^{-6}$ Euros   \\ \hline
Network resource share $T_{ij}$   &    20 ms   \\ \hline
Bandwidth $W$   &    20 Mhz   \\ \hline
Noise spectrum density $N_0$   &   -174 dBm    \\ \hline
Path loss attenuation $\sigma$          &  $3.6 * 10^{-6}$ \\ \hline
$Episodes$  & $8000$ \\ \hline
$Steps$     & $200$ \\ \hline
Discount factor $\gamma$     & $0.95$ \\ \hline
Replay buffer size $\mathcal{D}$  & 10000\\ \hline
Batch size $\mathcal{B}$  & $128$ \\ \hline
Optimizer    & Adam\cite{adam} \\ \hline
Critic learning rate $\eta_{\phi}$ & $0.0003$ \\ \hline
Actor learning rate $\eta_{\vartheta}$ & $0.0001$ \\ \hline
\end{tabular}
\end{table}
The simulations were conducted using the parameters as in Table \ref{table:parameters}. Moreover, to adjust the trade-off between the metrics according to the patient status, the hyperparameters $\alpha, \beta, \lambda$ and $\delta$ will have similar values when the patient has no seizure. However, $\lambda$ and $\delta$, which indicate the significance of the latency and distortion respectively, will have higher values than the others when the patient has a seizure. As for the seizure, we assume that the seizure can happen at any point in time with a probability $\upsilon_i$, where $\textbf{P}(\zeta^t_i = 1) = \upsilon_i$.

\subsection{Rewards Convergence and Policy Evaluation}

\begin{figure}[t!]
	\centering
	\includegraphics[scale=.7]{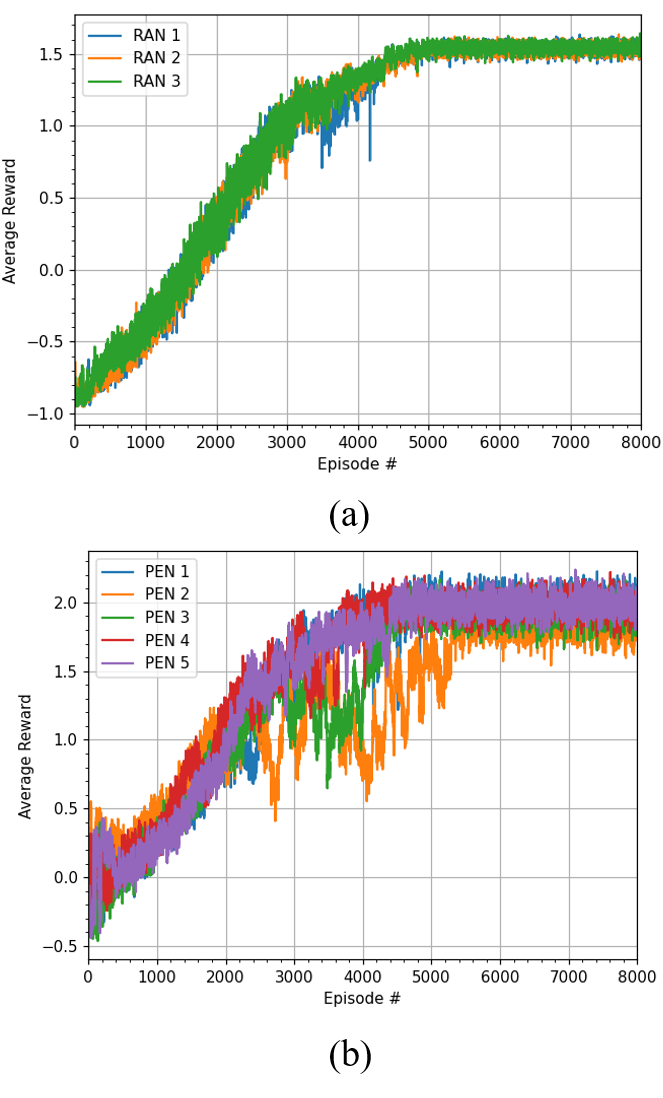}
	\caption{\footnotesize The achieved rewards during training for agents from (a) RAN team and (b) PEN team}
	\label{fig:rewards_convegence}
\end{figure}

In the training process of the agents, we ran Algorithm 1 for 6000 episodes, where in the first 500 episodes the agents' actions are fully exploratory. After that, the exploration starts decaying until the agents actions become fully exploitary. Fig \ref{fig:rewards_convegence} shows the achieved rewards for the agents during training. In Fig \ref{fig:rewards_convegence}(a), it can be seen that the RAN agents obtained the convergence around episode 4000. Similarly, in Fig \ref{fig:rewards_convegence}(b), most of the PEN agents obtained the convergence at episode 4000, except for \textit{PEN 2} agent, it converged around episode 5200.
\begin{figure*}
	\centering
	\includegraphics[scale=.6]{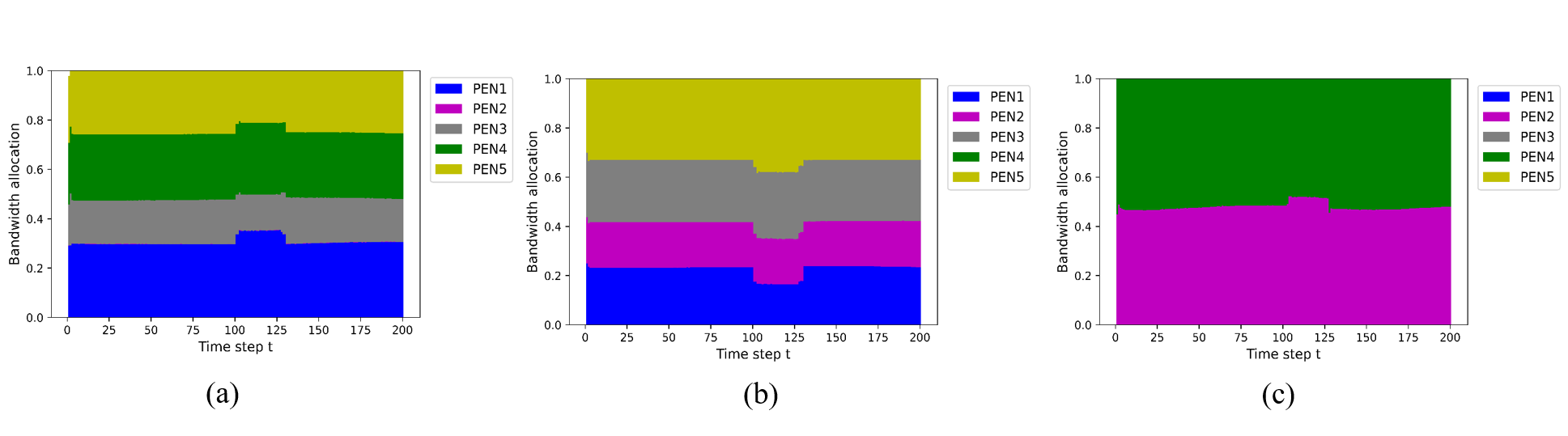}
	\caption{\footnotesize The learned policy for (a) \textit{RAN} 1 , (b) \textit{RAN} 2 and (c) \textit{RAN} 3 agents}
	\label{fig:Policy_RANs}
\end{figure*}

\begin{figure*}[!h]
	\centering
	\includegraphics[scale=.56]{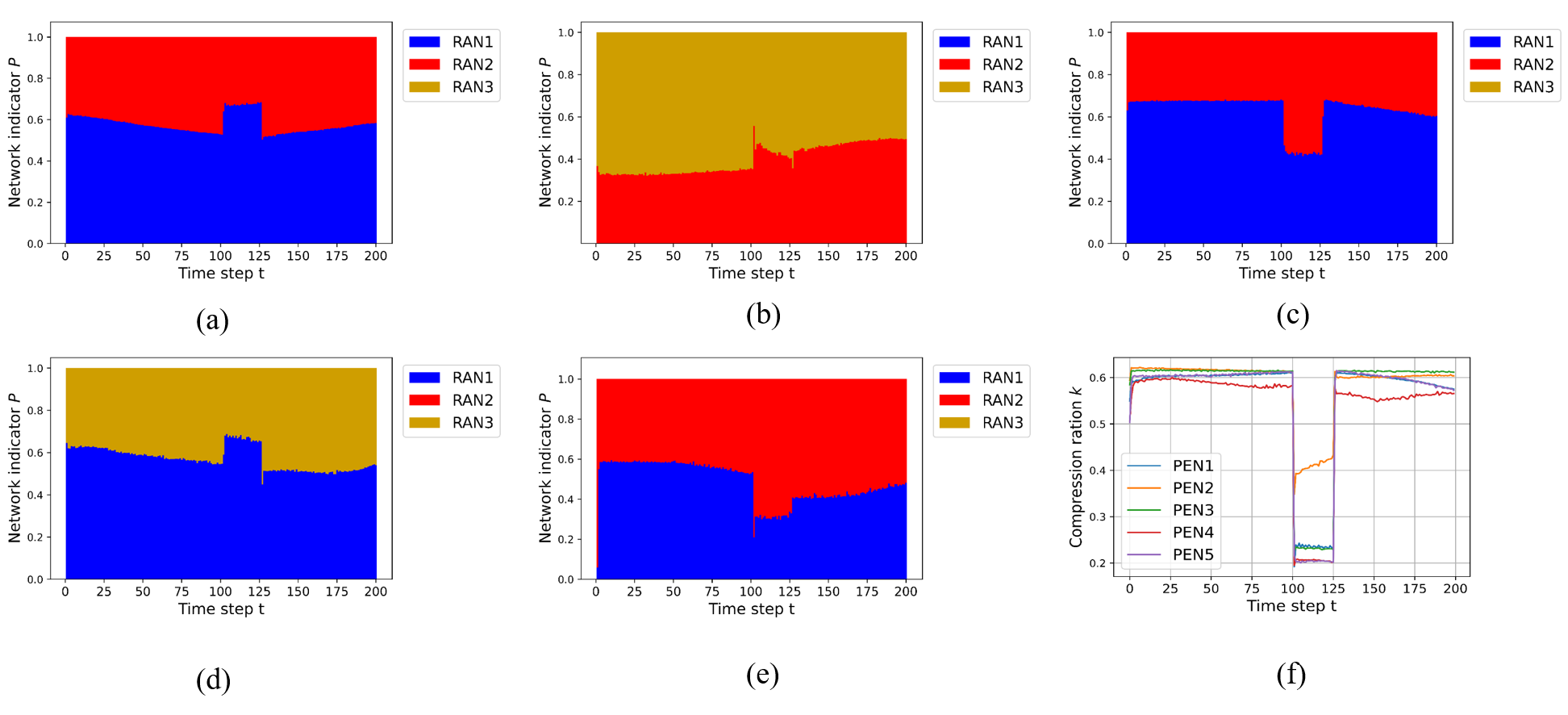}
	\caption{\footnotesize The learned policy for \textit{PEN} agents in terms of (a)-(e) network utilization, and (f) compression ratio}
	\label{fig:Policy_PENs}
\end{figure*}

\begin{figure*}
	\centering
	\includegraphics[scale=.52]{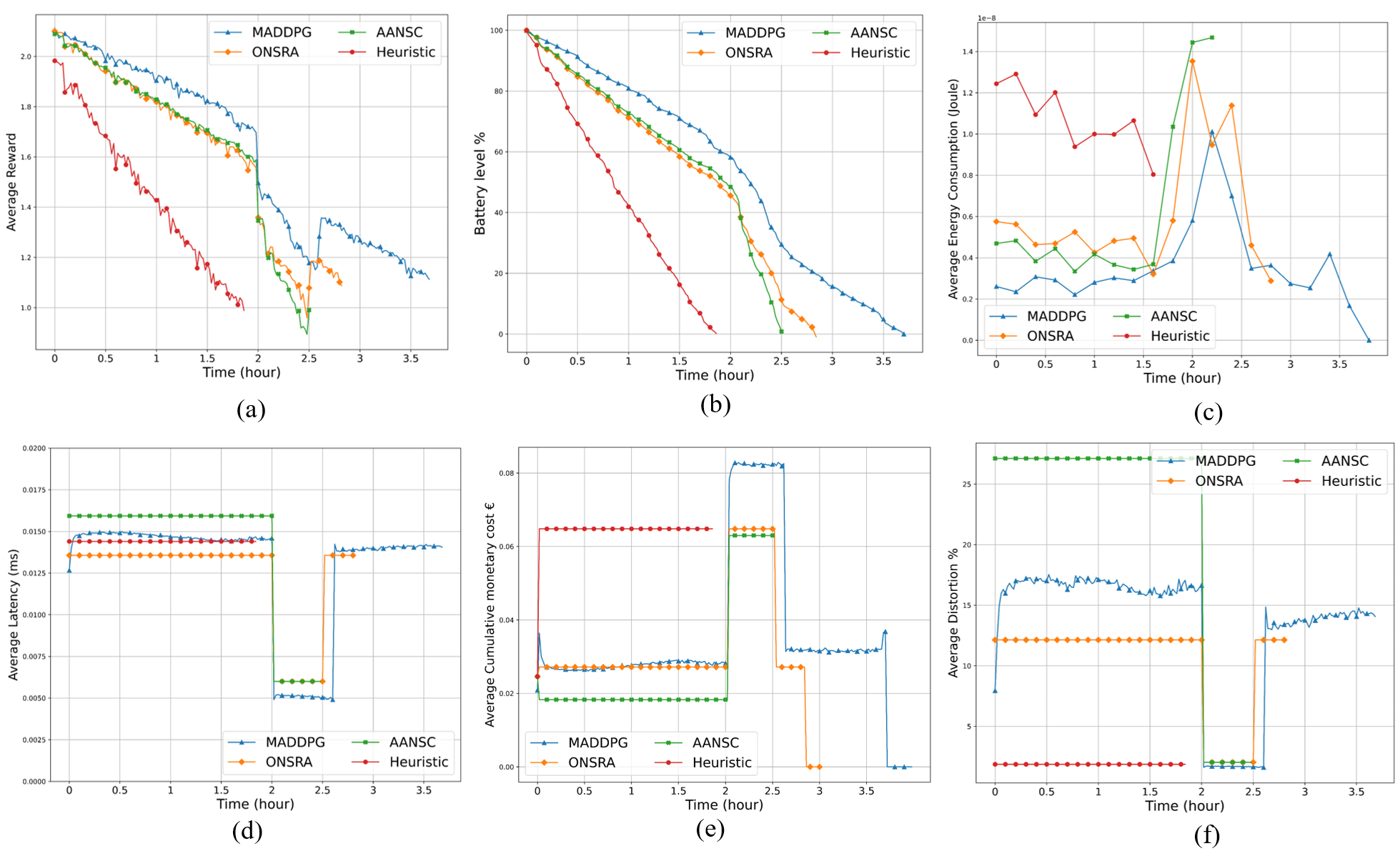}
	\caption{\footnotesize Comparison between our proposed and other algorithms in terms of average (a) obtained rewards, (b) battery lifetime, (c) energy consumption, (d) latency, (e) monetary cost and (f) distortion}
	\label{fig:comparison_all}
\end{figure*}
We show the learned policy for the agents from both teams, the \textit{PEN} and the \textit{RAN} teams during one episode of evaluation. The learned policy for the RAN team after training is depicted in Fig \ref{fig:Policy_RANs}. Firstly, in Fig \ref{fig:Policy_RANs}(a), we can see that \textit{RAN} 1, has the highest data rate and has allocated equal share of the bandwidth for \textit{PEN}s 1, 4 and 5 and a smaller bandwidth for \textit{PEN} 3, while \textit{PEN} 2 got neglected completely and has not been allocated anything by \textit{RAN} 1. Secondly, the \textit{RAN} 2 bandwidth allocation is shown in Fig \ref{fig:Policy_RANs}(b). It can be seen that \textit{PEN}s 1, 2, 3 have been allocated equal share of the bandwidth, while \textit{PEN} 5 has a slightly larger bandwidth, and \textit{PEN} 4 got neglected completely. Lastly, the bandwidth allocation for \textit{RAN} 3, which has the smallest data rate, is shown in Fig \ref{fig:Policy_RANs}(c). In fact, \textit{RAN} 3 could only allocate for only two \textit{PEN}s, which are \textit{PEN} 2 and 4, where allocating to more \textit{PEN}s will make the data rate for each one very low, and will not be sufficient to deliver the data on time.

The \textit{PEN}s agents learned policy is depicted in Fig \ref{fig:Policy_PENs}. The policy is represented by the network utilization indicators (i.e., $P_{ij}$'s) and the compression ratio.  In Fig \ref{fig:Policy_PENs}(a), we can see that \textit{PEN} 1  has utilized \textit{RAN} 1 the most, along with \textit{RAN} 2, while it did not send any data to \textit{RAN} 3 since it has no allocated bandwidth on it. As for \textit{PEN} 2, which its allocation is shown in Fig \ref{fig:Policy_PENs}(b), it had utilized \textit{RAN} 2 and 3 and ignored \textit{RAN} 1 as it has no bandwidth share on it. Similar to \textit{PEN} 1, \textit{PEN}s 3, and 5 have the same behavior as they utilize \textit{RAN} 1 and 2, while not sending any data to \textit{RAN} 3. Lastly, \textit{PEN} 4 had utilized \textit{RAN} 1 along with \textit{RAN} 3, and ignored \textit{RAN} 2. However, a noticeable pattern in the \textit{PEN}s policies exist, which is when the seizure happens, the \textit{PEN}s tend to utilize  the \textit{RAN}s more on which the \textit{PEN}s have more available bandwidth. Finally, the compression ratio for all the \textit{PEN}s is shown in Fig \ref{fig:Policy_PENs}(f). Interestingly, all the \textit{PEN}s maintain a high compression ratio, while at the seizure time, the compression drops down sharply to a very low value so that the medical data have the minimal possible distortion.

\subsection{Performance comparison}
In order to illustrate our proposed algorithm's performance, we compare it to multiple techniques, namely, a heuristic, Autonomous and adaptive Network Selection (AANSC) \cite{adap_comp_selection} and the Optimal Network Selection and Resource Allocation (ONSRA) algorithms \cite{ONSRA}. The heuristic algorithm will be the naive approach, which is the RANs allocate equal shares of the bandwidth to all the PENs, and the PENs to utilize all the RANs available equally, while setting the compression ratio to the minimum possible according to (\ref{eqn:Delay_const}). The AANSC algorithm assumes that all the PENs have equal shares of the bandwidth on all the RANs, while tries to optimize the energy consumption, monetary cost, latency and distortion at the PEN side. As for the ONSRA algorithm, it simplifies the problem $\textbf{P1}$ by sub-dividing it into two sub-problems. The first sub-problem is to optimize the resource allocation at the RANs side, given the network utilization indicators as constants from the PENs. Whereas the second sub-problem is the optimization of the energy consumption, monetary cost, latency and distortion at the PEN side after receiving the allocated bandwidth from the RANs, and it keeps exchanging the solutions of the two sub-problems between the RANs and the PENs until convergence. The heuristic approach will be the baseline, while the ONSRA will be the benchmark of the comparison. We show a sample for the comparison for one PEN in a setting, where we keep the PEN running for hours until the PEN runs out of battery. The comparison will be in terms of the average rewards achieved during the PEN lifetime according to \ref{eqn:reward_fun},  the battery lifetime, average energy consumption, latency, monetary cost and distortion, where Fig \ref{fig:comparison_all} depicts all the comparisons between our proposed approach and the aforementioned algorithms.

In Fig \ref{fig:comparison_all}(a), we can see the average obtained rewards during the PEN lifetime. The reward at first declines slowly as the battery level decreases over time. However, our proposed algorithm has a slower decline than the other approaches, and the heuristic has the fastest decline since it is the naive approach and does not take into consideration any optimization perspective. After two hours approximately, we can see the sharp drop in the rewards. This is due to the fact that the patient had a seizure at that time, and the reward function changed to signify the latency and the distortion importance while disregarding the energy consumption and the monetary cost. Moreover, the proposed approach had a lower drop than the other two approaches, the AANSC and the ONSRA, where the heuristic drained the battery before the seizure. The battery lifetime achieved by the algorithms is shown in Fig \ref{fig:comparison_all}(b). The heuristic approach has the worst battery lifetime, which is around 1.8 hours, while the AANSC and the ONSRA had achieved better PEN lifetimes with 2.5 and 2.8 hours respectively. Indeed, our proposed algorithm achieved the highest battery lifetime with 3.8 hours, where it achieved a lifetime which is better by 100\% more than the heuristic, and by 45\% and 30\%  more than the AANSC and the ONSRA respectively. In fact, the energy consumption of the algorithms, which is shown in \ref{fig:comparison_all}(c) justifies the aforementioned results, where our proposed approach had the lowest energy consumption on average. Moreover, when the patient had a seizure, we can notice a spike in the energy consumption, where it resulted in increasing the declining slop in the battery lifetime in the previous figure. 

As for the latency, which is depicted in Fig \ref{fig:comparison_all}(d), the algorithms achieved similar results, with a slight advantage to the ONSRA algorithm. However, at seizure time, our algorithm achieves a slightly better latency than the other algorithms. Afterwards, the cumulative monetary cost is shown in Fig \ref{fig:comparison_all}(e). While the heuristic had the highest cost, the other algorithms had similar overall cost. However, during the seizure, our algorithm had the highest overall cost. In fact, as mentioned before, the proposed algorithm disregards the trade-off between energy consumption, latency, cost and distortion, and focuses only on the latency and the distortion in order to minimize the medical data transmission time along with the minimum distortion possible. Finally, the distortion is shown Fig \ref{fig:comparison_all}(f). When the patient is not having a seizure, we can notice that the algorithms other than the heuristic allow some levels of distortion in order to minimize the energy consumption, latency and cost. Nevertheless, all the algorithms tend to have negligible distortion at seizure time.

\section{TB-MADDPG Complexity Analysis}\label{cmplx_analysis}
The TB-MADDPG algorithm employs neural networks to facilitate the agents' and critics' training, specifically, the multi-layer perceptron architecture (MLP). First, for a single-agent RL that employs the MLP architecture with three layers, it has been shown in \cite{PLS_complexity} that the training complexity is given by $O( e \times t (S\times \eta + \eta \times A) )$, where $e$ is the number of episodes of training, $t$ can be either the number of iteration per episode or the lifetime of a PEN. The $S$ denotes the input layer size, which also represents the agent's observations set size, $\eta$ is the hidden layer size, and $A$ denotes the output layer, which also represents the agent's actions set size. As for the critic, recall that in our algorithm, each critic in each team evaluate the joint actions and observations of the whole team with a single value. Thus, its training complexity is given by $O \big ( e \times t \big ( ( N\times (A+S) )\times \eta + \eta \times 1 \big ) \big )$, where $N$ is the number of agents in the team. Hence, the training complexity of one agent with its critic can be expressed as $O( \Omega )$, where $\Omega$ is given by:
\begin{equation}
\Omega = e \times t \bigg ( (S\times \eta + \eta \times A) + ( N\times (A+S) )\times \eta + \eta \times 1 \bigg )
\end{equation}
Consequently, the training complexity of one team with $N$ agents can by given by $O ( $N$ \, \Omega )$. Finally, since we have two teams, namely, the RANs and the PENs teams, with $M$ and $N$ agents in each team respectively, the TB-MADDPG training complexity is given by $O \big ( \Omega \, (N+M) \big )$. As for the execution complexity, since each agents acts on its own without the need for critic and other agents interactions, the complexity for taking one action at any given time step for each agent is the complexity of a single feed forward pass in a neural network, which is given by $O( S\times \eta + \eta \times A )$. 
\section{Conclusion}\label{sec:conclusion}
In this paper, we presented a novel approach for network selection along with adaptive compression at the PEN side, and resource allocation for the PENs at the RANs side. More precisely, we have proposed a DMARL algorithm, namely, the TB-MADDPG, which accounts for the heterogeneity of the agents. The TB-MADDPG groups each type of the agents in each team, and tries to find the optimal joint policy for the PENs team in terms of network utilization and adaptive compression, while finding the optimal policy for the RANs team in terms of bandwidth allocation to maximize the PENs QoE. The presented results in this paper shows that our proposed approach significantly outperforms the existing state-of-art techniques for network selection and resource allocation.  
\section*{Acknowledgement}
This work was made possible by NPRP grant \# NPRP12S-0305-190231 from the Qatar National Research Fund (a member of Qatar Foundation). The findings achieved herein are solely the responsibility of the authors.  

\balance

\bibliographystyle{IEEEtran}
\bibliography{references.bib}

\end{document}